\documentclass[12pt]{revtex4}%,draft,twocolumn,prl,showpacs,12pt
\usepackage{epsfig,color,amsfonts,amsmath,amsthm,comment,natbib}%,showkeys,natbib,
\newcommand{\ltwid}{\mathrel{\raise.3ex\hbox{$<$\kern-.75em\lower1ex\hbox{$\sim$}}}}
\newcommand{\gtwid}{\mathrel{\raise.3ex\hbox{$>$\kern-.75em\lower1ex\hbox{$\sim$}}}}

\def\nn{\nonumber\\}
\newcommand{\bea}{\begin{eqnarray}}
\newcommand{\eea}{\end{eqnarray}}
\theoremstyle{remark}
\newtheorem{remark}{Remark}

\begin{document}
\bibliographystyle{apsrev}
\title{A note on the Full Counting Statistics of paired fermions}

\author{Israel Klich}
\affiliation{Department of Physics, University of Virginia, Charlottesville, VA, USA}

%\date{\today}

\begin{abstract}
We study trace formulas for the exponentials of general fermion bilinears, including pairing terms, and including non Hermitian forms. In particular, we give elementary derivations for determinant and Pfaffian formulae for such traces, and use these to obtain general expressions for the full counting statistics in states associated with quadratic Hamiltonians, including BCS-like pairing terms and fermion parity in a prescribed region or set of modes. We also derive Pfaffian expressions for state overlaps and counting statistics in states built out of the vacuum by creation of pairs of particles.
\end{abstract}
\maketitle

\section{Introduction}
Fluctuations of observables in quantum mechanical systems carry important information about a variety of properties, such as the nature of transport mechanisms in mesoscopic systems or spin excitations in magnetic systems, and supply a tool for studying physical properties both in equilibrium and non-equilibrium settings. An important example for such is the use of shot noise, i.e. non-equilibrium current fluctuations, to get information about the nature of charge carriers in complicated interacting situations such as fractional quantum Hall systems, where collective excitations carry a fractional charge. Additional, more detailed information about condensed matter systems, may be hidden in higher order noise correlations. This information becomes accessible as our ability to carry out precision measurements increases. %Indeed, recent experiments have demonstrated how the full distribution of transmitted charges in certain systems may be obtained. 
The full distribution functions of transmitted charges, known as the "Full Counting Statistics" (FCS) has been used to describe transport in mesoscopic systems, and involve interesting problems both about the nature of transport and the nature of measurements themselves.

In the context of mesoscopic physics, the theory of FCS was pioneered in the celebrated work of Levitov and Lesovik \cite{levitov1992charge}, where a beautiful formula for the FCS of transport through a junction has been derived. Subsequently, FCS has seen intense work in mesoscopic physics \cite{levitov1992charge,beenakker2001counting,BlanterReview,nazarov2003quantum,nazarov2003full,schonhammer2007full,Esposito09} and in cold atom systems \cite{Polkovnikov06,Gritsev06}. The formulation of the FCS results for free fermions presented in \cite{Klich03}, has been successfully used to simplify FCS and related calculations and has proved useful also in other contexts, describing time dependent problems involving non-interacting fermions. For example, the FCS formalism has been found to play a useful role in the Fermi-edge problem \cite{muzykantskii2003fermi,abanin2005fermi}, and has been useful as a numerical tool for other time dependent problems, for example in studies of thermalization and decoherence of metallic leads \cite{kulkarni2013full}, as a tool to characterize correlations and phases \cite{ivanov2012characterizing}, identify dynamical phases in the  resonant-level model \cite{genway2012dynamical} as well as characterizing non-equilibrium situations such as the evolution of systems following a quench  \cite{eisler2013full}. FCS was studied theoretically and measured experimentally in several mesoscopic systems, such as tunnel junctions and quantum point contacts \cite{levitov2004counting,gershon2008detection,bomze2005measurement}
where non-gaussian fluctuations  \cite{gershon2008detection} and counting statistics of single electrons \cite{fujisawa2006bidirectional} were measured. Some attention has also been devoted to states where pairing terms appear in the Hamiltonian, to treat situations involving superconductivity. In such situations transitions between electron holes pairs and Cooper pairs via Andreev tunneling are present  (see, e.g.  \cite{muzykantskii1994quantum,belzig2001full,samuelsson2002proximity,braggio2011superconducting}). Recent experimental progress has been reported in the measurements of FCS of Andreev  tunneling in \cite{maisi2014full}. Shot noise signatures for systems with fractionalized charges have been proposed theoretically \cite{kane1994nonequilibrium,PhysRevLett.102.236402,milletari2013shot} and are used to experimentally access fractional charges \cite{de1997direct,saminadayar1997observation,inoue2013fractional}.
On the mathematical side, much work has been done to understand the thermodynamic limit of FCS on a more rigorous level, see e.g  \cite{de2007quantum,avron2008fredholm,derezinski2008fluctuations,jaksic2012entropic,bernard2012full}.

In a broader context, it has also been demonstrated that fluctuations, and the FCS of charge or other conserved quantities (such as, for example, block magnetization in certain spin chains), may contain information about the full entanglement scaling of a system split into two parts. It was shown how one can compute entanglement entropy and Renyi entropies from FCS for certain systems in \cite{klich2006measuring}, a measurement of entanglement entropy in a transport experiment using FCS was suggested in  \cite{klich2009quantum} and other systems were discussed in \cite{song2011entanglement,song2012bipartite}.  

%,calabrese2011entanglement

Many of the above advances hinge on the efficient calculation of partition function like objects - traces of exponents of quadratic fermion  Hamiltonians. However, paired Hamiltonians possess an additional layer of complexity compared to the non-interacting fermions, due to the lack of charge conservation. Algebraically, the lifting of single particle dynamics is non-unique, as the group of Bogolubov transformations is not simply connected, resulting in sign ambiguities. 

Here we present a rather simple closed formula, Eq. \eqref{Pfaff eq}, for the trace of a single exponent of quadratic fermion operators, where the sign ambiguity is not present. For products of exponents and most practical purposes, in the equations presented below, such as Eq. \eqref{general product of exp majorana},  the sign can be determined by simple analyticity arguments. In addition we supply simple formulas for full counting statistics and overlaps in pairing states as described in Eq. \eqref{overlap} and Eq. \eqref{counting in vacuum with pairing}.

\section{The trace of the exponential of a fermion bilinear}
In this section we show how to extend the well known formula for the partition function of a number conserving, bilinear, fermion Hamiltonian
\bea
 {{\rm{Tr}}}~  e^{-\beta H_{{ij}} a^{\dag}_i a_j}=\det  \left(1+e^{-\beta H}\right),
\eea
to situations where the quadratic form includes pairing terms.
As we will observe shortly, in working with Hamiltonians which include pairing terms it is convenient to represent the fermions using Majorana operators. To do so, consider an $n$-dimensional Hilbert space $\mathfrak{h}$ of possible single particle states (modes), and associated fermionic creation and annihilation operators $a^{\dag}_k, a_{k}$, with $k=1,..,n$, obeying the canonical anti-commutation relations: $a^{\dag}_ka_l+a_la^{\dag}_k=\delta_{lk}$ and $a_ka_l+a_la_k=0$.  We define $2n$ Majorana operators by
\bea&\label{majorana definition}
c_k=a_k^{\dag}+a_k~~;~~k=1,..,n~,\nn & ~~c_k=i (a_{k}^{\dag}-a_{k})~~;~~k=n+1,..,2 n~,
\eea
with the $c_k$ obeying the Clifford relations:
\bea\label{clifford}
c_k c_l+c_l c_k=2\delta_{kl}~.
\eea
To define taking traces below, we assume a standard  representation of the Clifford algebra on the $2^n$ dimensional Fock space associated with the fermions. In practice we use the algebraic relations, and fix the dimension of the representation to $2^n$ (alternatively, one may represent the algebra on a $2^{n}$ dimensional Hilbert space explicitly in terms of Pauli matrices, augmented by Jordan-Wigner strings).
 
In this section we derive the following results:

1) For a $2n$ dimensional, antisymmetric matrix $A\in Skew(2n,\mathbb{C})$, the following formula holds:
\begin{eqnarray}\label{Pfaff eq}
\mathcal{Z}(A)\equiv
 {{\rm{Tr}}}~  e^{A_{{ij}} c_i c_j}=\frac{{\rm Pf} \left(e^{-4 A}-e^{4 A}\right)}{{\rm Pf} \left(e^{-2 A}-e^{2 A}\right)}~.
\end{eqnarray}
were ${\rm Pf}$ is the Pfaffian, defined for a $2n\times 2n$ matrix $M$ by
\bea\label{det eq1}
{\rm Pf}(M)=\frac{1}{2^nn!}\Sigma _{\mathcal{P}\in \mathcal{S}_{2n}}
   (-1)^{\mathcal{P}}M_{P_1P_2}M_{P_2P_3}...M
   _{P_{2n-1}P_{2n}},
\eea
here $\mathcal{S}_{2n}$ is the permutation group on $2n$ elements. 

2) In addition, we note a simpler formula, which is correct up to a sign ambiguity:
\bea\label{det single ex}
\mathcal{Z}(A)=\sqrt{\det  \left(1+e^{4 A}\right)}.
\eea

3) Also, we note that the canonical mode decomposition available for Hermitian quadratic Hamiltonians, Eq. \eqref{canonical antisym}, may be extended to an arbitrary non-degenerate (possibly complex) antisymmetrix form $A_{{ij}} c_i c_j$ using a {\it similarity} transformation on Fock space, and into a Gantmacher mode decomposition, Eq. \eqref{Gantmacher Form}, when the degeneracy cannot be simply resolved.

Note the appearance of the square root in the determinant equation \eqref{det single ex}. In this equation, the sign of the determinant has to be determined. In many practical calculations the sign can be determined as follows.
Consider $\mathcal{Z}(\lambda A)$: from it's definition, Eq. \eqref{Pfaff eq}, it follows
 that $\mathcal{Z}(\lambda A)$ is an analytic function of $\lambda$. This determines the correct way of taking the sign of the square root: the sign has to be taken so that the right hand side of Eq. \eqref{det single ex} is everywhere analytic as well, and so that at $\lambda=0$ we have $\mathcal{Z}(0)=2^n$, the dimension of the corresponding Fock space. That also implies that any zeroes of $\det\left(1+e^{4 A}\right)$ must come in pairs, as to not create branch cuts in the complex $\lambda$ plane. In the next section we will repeat this type of argument with several other expressions involving square roots.

In particular, when $A$ is real antisymmetric, the equations \eqref{Pfaff eq} and \eqref{det single ex} are expressions for the partition function of paired fermions with the (hermitian) Hamiltonian $A_{{ij}} c_i c_j$, in such a case the square root in Eq. \eqref{det single ex} must be taken as positive. 

\section{Normal forms in Fock space and derivation of formulas}
To prove the above results, we need the following preliminary considerations.
Consider $O\in O(2n,\mathbb{C})$, a complex orthogonal transformation (i.e. $O^{T} O=1$). We emphasize that the group of complex orthogonal matrices is very different from the group of unitary matrices and the group of orthogonal matrices, in particular $ O(2n,\mathbb{C})$ is not compact. We note that for $O\in O(2n,\mathbb{C})$ the transformation:
\bea\label{trans1}
c_k\rightarrow O_{kl}c_l
\eea
preserves the anti-commutation relations \eqref{clifford}. 

Next, we establish the important fact that this transformation can always be written as a {\it similarity} transformation on the Clifford algebra obeyed by the $c_k$, i.e. there is some $X(O)$, and operator acting on Fock space, such that $X(O) c_k X(O)^{-1}=O_{kl}c_l$. To see this, note that any complex orthogonal transformation can be written as:
\bea
O=O_{R} ~e^{i K},
\eea
where $O_R$ is a real orthogonal matrix, $O_{R}\in O(2n,\mathbb{R})$ and $K\in Skew(2n,\mathbb{R})$ is a real antisymmetric matrix. 

The transformations $e^{i K}$ and $O_{R}\in SO(2n,\mathbb{R})$ can be generated on the $c_i$ by applying exponents of bilinears with antisymmetric form using 
\bea
(a)~~~c_m \Rightarrow e^{-{i\over 4}K_{ij}c_ic_j} c_{m} e^{{i\over 4}K_{kl}c_kc_l}=(e^{i K})_{ml}c_l,
\eea
for antisymmetric matrices $K\in Skew(2n,\mathbb{C})$. Transformations (a) are not enough, as these are restricted to exponents of antisymmetric matrices, which do not cover all complex orthogonal transformations (A simple example for an orthogonal matrix which cannot be written as an exponent of an antisymmetric matrix is $\sigma_z\in O(2,\mathbb{R})$).

To get the full $O(2n,\mathbb{R})$, we need to add similarity transformations that change signature. To generate such transformations we note that conjugation by $c_{k}$ is a similarity transformation since $c_{k}=c_{k}^{-1}$: \bea
(b)~~~c_{l} \Rightarrow c_{k} c_{l} c_{k}=-c_l+2\delta_{kl} c_k.
\eea
This transformation multiplies all the Majorana operators by $-1$ except for $c_k$ itself.
Together (a) and (b) transformations generate all possible $O\in O(2n,\mathbb{C})$.

{\it Proof of Eq  \eqref{Pfaff eq}.} 
Noting that under the transformation \eqref{trans1}, we have $A_{{ij}} c_i c_j \rightarrow  A_{{ij}} O_{i k}c_k O_{j l}c_l= (O^{T}AO)_{kl}c_kc_l$, we conclude that {\it for any } $O\in O(2n,\mathbb{C})$ the above similarity transformations preserve the trace, thus we can write:
\bea
\mathcal{Z}(A)=
 {\rm{Tr}}~  e^{A_{{ij}} c_i c_j}= {\rm{Tr}}~  X e^{A_{{ij}} c_i c_j} X^{-1}= {\rm{Tr}}~  e^{X (A_{{ij}} c_i c_j)X^{-1}} =\mathcal{Z}(O A O^T).
\eea

For A real antisymmetric, one may proceed  by transforming A into a canonical form. The eigenvalues of an antisymmetric matrix come in pairs $\pm \epsilon$, and using an (real) orthogonal transformation it can be written as:
\bea\label{canonical antisym}
A=O (\oplus_i E_i) O^{-1}~~~;~~~E_i= \left(
\begin{array}{cc}
 0 & \epsilon_i \\
   -\epsilon_i & 0 \end{array}
\right).
\eea
We note that for {\it real} symmetric matrices, this type of form is always available: it is essentially the Bogolubov transformation of the modes, and is the one often used when computing the partition function of a paired state. We can now write:
\bea
\mathcal{Z}(A)=
 {\rm{Tr}}~  e^{\sum \epsilon_i (c_{2i} c_{2i+1}-c_{2i+1} c_{2i})}=\prod_i 2\cos{2\epsilon_i},
\eea
where we used that: 
\bea
 {\rm{Tr}}~   e^{\epsilon(c_{2i} c_{2i+1}-c_{2i+1} c_{2i})}=2\cos{2\epsilon}.
\eea
Note that:
\bea
{\rm Pf} (e^{E_i}-e^{-E_i})={\rm Pf} (2\sin(\epsilon_i)i \sigma_y)=2\sin(\epsilon_i),
\eea
which gives
\bea
(2\cos{2\epsilon_i})={\sin(4 \epsilon_i)\over \sin(2 \epsilon_i)}={{\rm Pf} (e^{4 E_i}-e^{-4 E_i})\over {\rm Pf} (e^{2 E_i}-e^{-2E_i})}.
\eea
Combining these results together, and using that both Pfaffians and determinants of block matrices are products, we have Eq. \eqref{Pfaff eq}. Alternatively,  we can write
$$
(2\cos{2\epsilon_i})^2=(1+e^{4 i \epsilon_i})(1+e^{{-4 i \epsilon_i}})=\det(1+e^{4 E_{i}}), $$
giving us Eq. \eqref{det single ex}.

We now show that the form  \eqref{Pfaff eq} holds also for arbitrary complex antisymmetric matrices $A$, in order to be able to account for complex phases as may appear in mean field superconducting Hamiltonians as well as products of exponents of bilinears.

We claim the formula still holds. Indeed, it follows from the definition of $\mathcal{Z}(A)$, that $\mathcal{Z}(A)$ is manifestly an entire function of the elements  $A_{ij}$. On the other hand the Pfaffian expression in $\eqref{Pfaff eq}$ is also entire. To see this note that as a ratio of holomorphic functions, the only potentially problematic points are the points where the denominator goes to zero faster than the numerator, making the ratio singular. However, since we have (using 
${\rm Pf} (M)^2=\det(M)$) the relation:
$$
\Big[\frac{{\rm Pf} \left(e^{-4 A}-e^{4 A}\right)}{{\rm Pf} \left(e^{-2 A}-e^{2 A}\right)}\Big]^2={\det(e^{-4 A}-e^{4 A})\over \det(e^{-2 A}-e^{2 A})}=\det(e^{-2A}+e^{2A}),
$$
where the right side is manifestly regular, we conclude that this type of singularity can not happen. By analyticity, we conclude that the identity \eqref{Pfaff eq} must hold in the complex case as well.

In fact, it is possible to make a stronger argument, in that a canonical form is available even in the case of complex matrices, albeit as a {\it similarity} transformation on the Majorana modes, rather than an orthogonal transformation.
Indeed, we consider a Gantmacher type decomposition, described in the classic book \cite{gantmacherapplications}. This decomposition supplies a complex orthogonal diagonalization of the matrix which is rather complicated: 
Let $A$ be a rank $r$ anti-symmetric matrix, with elementary divisors $\epsilon_i$ and corresponding ranks $f_i$, then there exists a matrix $O\in O(2n,\mathbb{C})$ such that 
\bea
A=O K O^{-1},
\eea
where:
\bea
K=\oplus K_{\epsilon_i},
\eea
the $K$ matrices are of size $2f_i$, and they are of the form% $K_{\epsilon_i}=\kappa-\kappa^T$
%\bea&\label{Gantmacher Form}
%(K_{\epsilon_i})_{lm}={1\over 2}\{\sum_{l=1}^{f_i}(\delta_{m,l+1}+2 \epsilon_i \delta_{m,2f_i-l+1}-\delta_{f_i+m,f_i+l+1}\\ \nonumber & +i (\delta_{m,2f_i-l}+\delta_{m+1,2f_i-l+1})\}- (Transpose).
%\eea
%where
\bea&\label{Gantmacher Form}
K_{\epsilon}=\kappa-\kappa^T \\ &\nonumber
\kappa_{lm}={1\over 2}\begin{cases}  -\delta_{l,m+1}  &  m,l<f   \\
-i \delta_{m,f-l}-2\epsilon \delta_{m,f-l+1}-i \delta_{m,f-l+2}   & l>f~and~m<f    \\
 \delta_{l,m+1} & m,l>f     \end{cases}
\eea
For example, for $f=4$ the form is, explicitly:
\bea
{1\over 2}\left(
\begin{array}{cccccccc}
 0 & 1 & 0 & 0 & 0 & 0 & i & 2 \epsilon  \\
 -1 & 0 & 1 & 0 & 0 & i & 2 \epsilon  & i \\
 0 & -1 & 0 & 1 & i & 2 \epsilon  & i & 0 \\
 0 & 0 & -1 & 0 & 2 \epsilon  & i & 0 & 0 \\
 0 & 0 & -i & -2 \epsilon  & 0 & -1 & 0 & 0 \\
 0 & -i & -2 \epsilon  & -i & 1 & 0 & -1 & 0 \\
 -i & -2 \epsilon  & -i & 0 & 0 & 1 & 0 & -1 \\
 -2 \epsilon  & -i & 0 & 0 & 0 & 0 & 1 & 0 \\
\end{array}
\right)
\eea

Since we have shown above that all $O\in O(2n,\mathbb{C})$ transformations can be lifted as similarity transformations to the Fock space, we can bring any $A$ to this form.
Next, instead of treating the general Gantmacher form in full generality, we can proceed by first assuming that characteristic numbers are non degenerate. In this case we have $f_i=1$ for all $\epsilon_i$, the Gantmacher form reduces to the form \eqref{canonical antisym}, and one arrives at Eq. \eqref{Pfaff eq} directly.
We may also remove the condition on non-degeneracy of the eigenvalues of the matrix, by invoking the analyticity argument again. Starting with a degenerate anti-symmetric matrix, we can perturb it with an arbitrarily small deformation into a non-degenerate matrix, where the equality has been established. 

\begin{remark}
We note that the appearance of square roots is a consequence of the nature of the Majorana representation. The topological reason for this is that the Clifford representation of the Lie algebra of skew-symmetric matrices, even for just one fermion mode (two Majoranas) when exponentiated, corresponds to a double cover. This can be seen in the following simple example: %, with one (Dirac) fermion state, associated with the two Majorana operators $c_1,c_2$. 
Take $A=0$ and $B_{kl}=2\pi i (\sigma_y)_{kl}$ if $k,l\in \{1,2\}$ and $B_{kl}=0$ otherwise. Noting that the eigenvalues of $B$ are just $\pm 2\pi i$, we immediately have:
\bea
e^A=e^B={\bf I}_{2n}, 
\eea
and
\bea
e^{{1\over 4}A_{kl}c_kc_l}=e^{0}={\bf I}_{2^{n}},
\eea
where ${\bf I}_{2n},{\bf I}_{2^{n}}$ are the identity matrices in the $2n$ dimensional Majorana mode space and $2^{n}$ dimensional Fock space, respectively.
On the other hand:
\bea
e^{{1\over 4}B_{kl}c_k c_l}=e^{{\pi\over 2}(c_1c_2-c_2c_1)}\neq {\bf I}_{2^{n}}.
\eea
For example, taking $n=1$, we have:
\bea
{\rm Tr} e^{{\pi\over 2}(c_1c_2-c_2c_1)}=-2=-{\rm Tr} 1.
\eea
So when writing ${\rm Tr} e^{{1\over 4}A_{kl}c_k c_l}$ in terms of properties of $e^A$, the information about the sign comes in a subtle way. This is exactly what our Pfaffian formula \eqref{Pfaff eq} keeps track of. Indeed, computing the same expression using the Pfaffian we have
\bea
 {\rm{Tr}}~  e^{ \frac{\pi
   }{2}\left(c_1c_2-c_2c_1\right)}=\frac{\text{Pf
   }\left(e^{-2\pi  i \sigma _y}-e^{2\pi  i
   \sigma _y}\right)}{\text{Pf}\left(e^{-\pi  i
   \sigma _y}-e^{\pi  i \sigma
   _y}\right)}=\frac{\text{Pf}(0)}{\text{Pf}(0)}.
\eea
To resolve the ratio, we compute the same expression as a limit:
\bea &
\lim _{\epsilon \to 0} {\rm{Tr}}~  e^{ \frac{\pi
   +\epsilon }{2}\left(c_1c_2-c_2c_1\right)}=\lim
   _{\epsilon \to
   0}\frac{\text{Pf}\left(e^{-2(\pi +\epsilon ) i
   \sigma _y}-e^{2(\pi +\epsilon ) i \sigma
   _y}\right)}{\text{Pf}\left(e^{-(\pi +\epsilon
   ) i \sigma _y}-e^{(\pi +\epsilon ) i \sigma
   _y}\right)}=\lim _{\epsilon \to
   0}\frac{\text{Pf}\left(-4 \epsilon  i \sigma
   _y\right)}{\text{Pf}\left(2 \epsilon  i \sigma
   _y\right)}=-2.
\eea

\end{remark}

\section{The Trace of a Product of exponents}
For applications it is often more important to understand how to extend the previous results to deal with products of exponents, for example in order to compute expectation values. 
For this case we derive the following trace formula:
\bea&\label{general product of exp majorana}
 {\rm{Tr}}~  ~ e^{A_{1ij}c_i c_j}...e^{A_{nij}c_i c_j}=\sqrt{\det(1+e^{4A_1}..e^{4A_n})}~.
\eea
Finally, the expectation value of a product of exponents, in a thermal state with a Hamiltonian $H_{ij}c_{k}c_{l}$ is:
\bea&\label{general expectation Majorana}
%{ {\rm{Tr}}~  ~e^{-\beta H_{ij}c_i c_j} e^{(A_1)_{ij}c_i c_j}...e^{A_{nij}c_i c_j}\over  {\rm{Tr}}~  ~ e^{-\beta H_{ij}c_i c_j}}=
\langle e^{A_{1ij}c_i c_j}...e^{A_{nij}c_i c_j}\rangle=\sqrt{\det(n_{\beta}+(1-n_{\beta})e^{4 A_1}..e^{4A_n})}~, \\ & n_{\beta}={(1+e^{4 \beta H})}^{-1}.
\eea
To derive Eq. \eqref{general product of exp majorana} we first note that:
\bea
\left[\frac{1}{4}K_{lm
   }c_lc_m,\frac{1}{4}L_{ij}c_i
   c_j\right]=\frac{1}{4}( [K,L])_{i
   m}c_{i }c_m,
\eea
which is straightforward to check (see, e.g. \cite{kitaev2006anyons}).
Showing that the map:
\bea
K\rightarrow  {1\over 4}K_{lm
   }c_lc_m
\eea
is a representation of the Lie algebra of skew-symmetric matrices. In particular it follows that for small enough $|t|<t_{0}$, we can write:
\bea
e^{{t\over 4}A_{lm
   }c_lc_m}e^{{t\over 4}B_{lm
   }c_lc_m}=e^{{1\over 4}C[t A,t B]_{lm
   }c_lc_m} \rightarrow  e^{t A}e^{t B}=e^{C[t A,t B]},
\eea
where $C[t A,t B]$ is given by a Baker Campbell Hausdorff (BCH) type series. We kept $t$ since in the BCH formula and it's explicit variants such as the Dynkin formula, $C[tA,tB]$ as a series in commutators of $A$ and $B$, has in general a finite radius of convergence (see remark 2 bellow). In this neighborhood, we can immediately write: 
\bea&
 {\rm{Tr}}~ e^{{t\over 4}A_{ij}c_i c_j}e^{{t\over 4}B_{ij}c_i c_j}= {\rm{Tr}}~ e^{{1\over 4}C_{ij}c_i c_j}=\sqrt{\det(1+e^{C})}=\sqrt{\det(1+e^{t A}e^{t B})}~.
\eea
Now we invoke analyticity again: the LHS is an entire function of $t$, and in particular it's square is. On the other hand $\det(1+e^{t A}e^{t B})$ is also an entire function, and since the functions are equal for $t<t_{0}$, they are equal everywhere. It is left to take the square root and resolve the sign so as to be an entire function, which goes to $2^n$ at $t=0$. Finally, we establish \eqref{general product of exp majorana} by renaming $\{A, B\}\rightarrow \{4A,4B\}$, and repeating the argument iteratively for an arbitrary number of matrices. Unfortunately, the Pfaffian expression Eq. \eqref{Pfaff eq}, is not available in a simple form anymore, since there is no simple way of expressing the denominator, which requires the square root of $e^{C}$.

\begin{remark}
The question of the range of analyticity of the BCH formula has been studied in many works since the classic paper of Wei \cite{wei1963note}.
The possible non-analyticity of BCH can be demonstrated with an example. Following the approach presented in \cite{wei1963note}, we construct such an example for our particular Lie algebra of complex skew-symmetric matrices by searching for a pair of  such matrices where 
$[A,B]=A$. The smallest non-commuting algebra of skew-symmetric matrices is of $Skew(3,\mathbb{C})$. Consider the following pair of anti-symmetric matrices:
\bea
A=\left(
\begin{array}{ccc}
 0 & 1 & 1 \\
 -1 & 0 & -i \sqrt{2} \\
 -1 & i \sqrt{2} & 0 \\
\end{array}
\right)\text{     };\text{    }B=\left(
\begin{array}{ccc}
 0 & 1-i \sqrt{2} & 1 \\
 -1+i \sqrt{2} & 0 & -1-i \sqrt{2} \\
 -1 & 1+i \sqrt{2} & 0 \\
\end{array}
\right).
\eea
An explicit computation shows that for any scalars $a,b$:
\bea
e^{a(-2B)}e^{b A}=e^{\Big(a (-2B)+\frac{2 a b e^{2 a}}{e^{2 a}-1}
   A\Big)},
\eea
so  that
\bea
C[-2 a B,b A]=a (-2B)+\frac{2 a b e^{2 a}}{e^{2 a}-1}
   A.
\eea
Clearly, the BCH $C$ is not analytic in the input matrices $A,B$ close to $a={\pi i}$.

To go around this issue, one possible attempt is to use upper triangular matrices $A,B$ instead of antisymmetric ones, since $A_{ij}c_ic_j=2A^{up}_{ij}c_ic_j$ where $A^{up}$ is the upper triangular part of $A$.

The algebra of upper triangular matrices is nilpotent, and as such all BCH type series terminate and converge. However, it is straightforward to check that the map $A^{up}\rightarrow A^{up}_{ij}c_ic_j$ is not a representation of the Lie algebra of upper triangular matrices. We do note that upper-triangular matrices $A^{up}$, can be very useful in obtaining various relations between fermionic traces and determinants or Pfaffians. An example for such is Lieb's theorem on Pfaffians \cite{lieb1968theorem}. An alternative derivation of \eqref{Pfaff eq} in a manner similar to Lieb's derivation may be possible.

\end{remark}

\section{Applications to Full counting statistics}
Here, we consider counting statistics in systems evolving with a time dependent mean field BCS type hamiltonian as represented by Bogolubov-de Gennes equations. %For simplicity, we assume that the measured charge at time $t=t_0$ is a good quantum number. More precisely,
Consider measuring an observable $Q_A$ at time $t=t_A$, and measuring an observable $Q_B$ at a later time $t=t_B$. The usual two-measurement protocol for counting statistics of the difference between the measurement of $Q_A$   and $Q_B$   is conveniently described by the cumulant generating function:
\begin{eqnarray}\label{counting stat gen}&
\chi(\lambda)=\sum_{a,b} p_a~ prob(a\rightarrow b) e^{-\lambda (Q_A(a)-Q_B(b))},
\end{eqnarray}
where $p_a=\langle a| \rho|a\rangle$ is the probability to be in state  $|a\rangle$ at the initial time $t=t_A$, and $\rho$ is the initial density matrix.

From now on, we take $Q_A=Q_B\equiv Q$, which is the case appearing in most applications (such as charge transport), but the extension to $Q_A\neq Q_B$ is straightforward.
If we take $|a\rangle$ to be a complete set of eigenstates of a (hermitian) operator $ \hat{Q}$ with eigenvalues $\hat{Q}|a\rangle=Q(a)|a\rangle$ representing possible outcomes of the observable $Q$,  we can write Eq. \eqref{counting stat gen} as 
\begin{eqnarray}
 \chi(\lambda)= {\Sigma_a }p_a\langle a| U^\dag e^{\lambda  \hat{Q}}U e^{-\lambda  \hat{Q}}|a\rangle,
\end{eqnarray}
where $U$ is the evolution operator of the full many body system from $t_A$ to $t_B$. 

In the simplest case, the measured states $|a\rangle$ are also eigenstates of the initial $\rho$. 
For example, in the case of charge measurements, we take $[\rho,{\hat Q}]=0$. An example for such a situation is a normal lead that is connected at time $t_A$ with a superconductor, or where pairing is turned on at time $t_A$. In this case Eq. \eqref{counting stat gen} further simplifies to:
\begin{eqnarray}&
\chi(\lambda)={\rm Tr}~\rho~ U^\dag e^{\lambda  \hat{Q}}U e^{-\lambda  \hat{Q}}=\langle   U^\dag e^{\lambda  \hat{Q}}U e^{-\lambda  \hat{Q}} \rangle.
\end{eqnarray}

To connect with the formalism above, consider fermions in a Fock space built from the single particle Hilbert space $\mathfrak{h}$. We first express bilinear fermion forms in terms of majoranas and then use Eq. \eqref{general product of exp majorana}.
Indeed, consider a general bilinear fermion operator as:
\begin{eqnarray}\label{fermion bilinears}
\mathcal{M}=\sum_{ij} M_{{ij}} a_i^\dag a_j +\frac{1}{2}\left\{\Delta _{-ij}  a_i a_j +\Delta _{+ij} a_i^\dag a_j^\dag \right\}~,
\end{eqnarray}
where the matrices $\Delta _{-},\Delta _{+}$ are assumed to be {\it anti-symmetric} (a choice that can  always be made). Note that, in general, we do not demand that the operator $\mathcal{M}$ is hermitian, hence we do not assume, a priori, a conjugacy relation between $\Delta_{-}$ and $\Delta_{+}$, nor do we assume $M$ to be Hermitian. We rewrite the operator in terms of majoranas as:
\begin{eqnarray}\label{fermion to majorana bilinears}
\mathcal{M}={1\over 4}\mathbb{M}_{ij}c_{k}c_{l}+{1\over 2}{\rm Tr} M
\end{eqnarray}
with
\bea&
\mathbb{M}= \left(\!
\begin{array}{cc}
 M_{a}\!+\!\frac{1}{2}  \left(\! \Delta _- +\Delta _+ \! \right) & i M_{s} +\frac{i}{2} \left(\! \Delta _- -\Delta _+ \! \right) \\
 -i M_{s} \!+\!\frac{i}{2} \left(\! \Delta _- -\Delta _+ \! \right) & M_{a} \!-\!\frac{1}{2}  \left(\! \Delta _- +\Delta _+ \!\right) \\
\end{array}
\! \right),\nonumber
\eea
where $M_{s/a}={1\over 2}(M\pm M^T)$ are the symmetric and antisymmetric parts of $M$ and $\mathbb{M}$ acts on the space $\mathfrak{h}\otimes {\mathbb{C}}^{2}$.
We then have, by Eq. \eqref{general product of exp majorana},
\bea\label{trace of several}&
 {\rm{Tr}}~\Pi_m e^{\alpha _m
   \mathcal{M}_m}=e^{{1\over 2}\sum_m {\rm Tr}~M_m}\sqrt{\det  \left(1+\Pi _m
   e^{\alpha _m
   \mathbb{M}_m}\right)}.
%\nn & \phi={1\over 2}\sum_m {\rm Tr} K_m
\eea

%where in the last line the trace is expressed as a determinant, up to a sign which has to be determined separately.
When the evolution is governed by a Hamiltonian $\mathcal{H}$, we can write:
\begin{eqnarray}\label{MainFormula}&
\chi(\lambda)=\langle e^{+i t \mathcal{H} } e^{\lambda  \hat{Q}}e^{-i t \mathcal{H} } e^{-\lambda  \hat{Q}}\rangle_{\beta}=\nn & \text{det}^{1/2}
   \left(n_\beta+\left(1-n_\beta\right) e^{
   i t \mathbb{H}}e^{\lambda \mathbb{Q}}e^{- i t\mathbb{H}}e^{-\lambda \mathbb{Q}}\right)
\eea
with:
\bea
n_{\beta}=\frac{1}{1+e^{\beta\mathbb{H}_0}}
\eea
playing the role of the Fermi function of the initial Hamiltonian $\mathcal{H}_{0}$. For a Hermitian operator $Q$, we see that for real $\lambda$, we must also have $\chi(\lambda)>0$. Therefore, in taking the square root, we take the branch of $\chi(\lambda)$ which is real and positive everywhere on the real $\lambda$ axis.

\begin{remark}
In the case where the initial state is not diagonal in eigenstates of the charge operator considered, one has to replace the initial density matrix ${\rho}$ by a ``decohered''  initial density matrix in the charge basis $\tilde{\rho}$:
\bea
\rho\rightarrow \tilde{\rho}=\sum_a  \langle a|\rho|a \rangle    |a \rangle \langle a|
\eea
which can be, for example, implemented by an auxiliary integral.
\end{remark}

\begin{comment}
As a corollary, we have the following formula. Consider the thermal expectation value of a product of bilinears in a thermal state defined by a bilinear Hermitian hamiltonian $\mathcal{H}$. associated with $\mathcal{H}$ is a matrix $ \mathbb{M}_0$ given by:
\bea\label{BCS ham M}
 \mathbb{M}_0=\left(
\begin{array}{cc}
 \text{Im} (\Delta +H) & \text{Re} (\Delta +H) \\
 \text{Re} (\Delta -H) & \text{Im} (H-\Delta ) \\
\end{array}
\right)
\eea
And we have:
\bea\label{MainFormula}&
\left\langle \Pi _m e^{\alpha _m
   \mathcal{M}_m}\right\rangle
   =\frac{ {\rm{Tr}}~  e^{-\beta  \mathcal{H}}
   \Pi_{m=1}^N e^{\alpha _m
   \mathcal{M}_m}}{\mathcal{Z}}=\nn & e^{i\phi}\text{det}^{1/2}
   \left(n_\beta+\left(1-n_\beta\right)  \Pi_{m=1}^N e^{i
   \alpha _m \mathbb{M}_m}\right)
\eea
with:
\bea
n_{\beta}=\frac{1}{1+e^{-i \beta\mathbb{M}_0}}
\eea
playing the role of the Fermi function.
\end{comment}

\section{Full counting statistics of the number of fermions in a region}
For certain applications one does not need two measurements, for example when considering measuring the full counting statistics of a given observable. In such a case we can extract the probability distribution from an expression of the form:
\bea
\chi(\lambda)=\sum_{a} p_a e^{\lambda Q_A(a)}=\langle e^{\lambda Q}\rangle.
\eea
As an example for such a situation, let us consider the counting statistics of the number of particles in a region $A$ of a thermal state defined by a BCS type Hamiltonian:
\bea
\mathcal{H}=\sum_{ij} H_{{ij}} a_i^\dag a_j +\frac{1}{2}\Delta_{ij}  a_i a_j +h.c. 
\eea

Using the hermiticity of $\cal H$ and \eqref{fermion to majorana bilinears} we associate with $\mathcal{H}$ the matrix $ \mathbb{H}_0$ given by:
\bea\label{BCS ham M}
 \mathbb{H}_0=\left(
\begin{array}{cc}
i \text{Im} (H+\Delta) & i \text{Re} (\Delta +H) \\
 i \text{Re} (\Delta -H) & i \text{Im} (H-\Delta ) \\
\end{array}
\right)
\eea
Noting that the number operator in region $A$ can be written, using Eq. \eqref{fermion to majorana bilinears}, as
\bea
\hat{N}_A=\sum_{i\in A}  a_i^\dag a_i=\sum_{i,j}  {P_A}_{i,j} a_i^\dag a_{j}=-\sum_{i,j} {1\over 4}(P_A\otimes \sigma_y)_{ij}c_ic_j+{1\over 2}{\rm Tr}\, P_A
\eea
where $ {P_A}_{i,j}=\delta_{ij}\Theta(i\in A)$ are the matrix elements of $P_A$, the single particle projector on region $A$. We have
\bea&
\chi_A(\lambda)\equiv \frac{ {\rm{Tr}}\, e^{-\beta \mathcal{H}_{0}} e^{\lambda \hat{N}_A}}{\mathcal{Z}}=\nn & e^{\frac{1}{2}
   \lambda   {\rm{Tr}}\,P_A} \!\sqrt{\det 
   \left(\!\frac{1}{1+e^{\beta 
   \mathbb{H}_0}}+\!\left(1-\frac{1}{1+e^{ \beta 
   \mathbb{H}_0}}\right) e^{-\lambda 
   P_A\otimes \sigma_{y}}\!\right)}\label{NumberCountingStatisticsBCS}
\eea 
where $\sigma _y$ acts on the auxiliary $\mathbb{C}^{2}$ space, so 
$$ P_A\otimes \sigma_{y}=
\left(
\begin{array}{cc}
 0 &  i P_{A}\\
-i P_{A}  &  0 
\end{array}
\right).$$
We note that here the sign of the square root of the determinant is unambiguous: as before, the sign of $\chi_A(\lambda)$ is assured to be positive for real $\lambda$ by it's definition, Eq. \eqref{NumberCountingStatisticsBCS}.  

As a an example, for simplicity, we consider a BCS type Hamiltonian \eqref{BCS ham M}, and we simply take $A$ to be the {\it entire} volume of the system, allowing us to work in momentum space. Written in a momentum state basis we will take $H=\sum_{p}h(p) a_{p}^{\dag}a_{p}$, where $h(p)$ may, for example, be taken the form $h(p)=\frac{p^2}{2 m}-\mu$. Also, we take a simple $s$-wave superconducting gap $\Delta$, only pairing states with momenta $p$ and $-p$. In this case we have to consider each pair of $p,-p$ separately. For each paired couple with a given $p$, we must consider a $4\times 4$ block from $ \mathbb{H}_0$ of the form:
\bea
\left(
\begin{array}{cccc}
 0 & 0 & h(p) & -\Delta(p)  \\
 0 & 0 & \Delta(p)  & h(p) \\
 -h(p) & -\Delta(p)  & 0 & 0 \\
 \Delta(p)  & -h(p) & 0 & 0 \\
\end{array}
\right)~.\label{MforUnifromBCS}
\eea
Plugging the matrix \eqref{MforUnifromBCS} into 
\eqref{NumberCountingStatisticsBCS}, and doing some algebra we find:
\bea&
\chi(\lambda,p) =\frac{e^{\lambda}}{2} 
   \text{sech}^2\left(\frac{\beta}{2}  
   \sqrt{h^2+\Delta ^2}\right) \times \label{counting stat for one pair}\\ &\left(1+\cosh
   \left(\beta  \sqrt{h^2+\Delta ^2}\right)
   \cosh (\lambda )-\frac{h \sinh \left(\beta 
   \sqrt{h^2+\Delta ^2}\right) \sinh (\lambda
   )}{\sqrt{h^2+\Delta ^2}}\right),\nonumber
\eea
and finally:
\bea &
\log(\chi(\lambda))=\int g(p)dp \Big\{\lambda +\log \big(1+\nn & \cosh \left(\beta \sqrt{h^2+\Delta ^2}\right) \cosh(\lambda)-\frac{h \sinh\left(\beta  \sqrt{h^2+\Delta^2}\right) \sinh (\lambda)}{\sqrt{h^2+\Delta^2}}\big)-\nn & \log  \left(2 \cosh^2\left(\frac{1}{2} \beta \sqrt{h^2+\Delta^2}\right)\right)\Big\},
\eea
where $g(p)$ is the density of pairs at a given $p$. 

To verify our result, we can also derive this formula directly. Indeed, write the pairing Hamiltonian (suppressing the momentum index $p$),
\bea
\mathcal{H}=h (a^\dag a+ b^\dag b)+\Delta  \left( a^\dag b^\dag+b a\right).\eea
Affecting a Bogolubov transformation we can write $\mathcal{H}$ in the form
\bea &
\mathcal{H}=\mathcal{E} c^\dag
   c-\mathcal{E} d^\dag d~~~~;~~~~\mathcal{E}=\sqrt{h^2+\Delta ^2},
\eea
with
\bea& c=u a+v b^\dag~~;~~d=v a-u
b^\dag,
 \eea 
where \bea u=\cos\!
\left(\!\frac{1}{2} \tan
^{-1}\!\left(\!\frac{\Delta}{h}\!\right)\!\right)~;~v=\sin\!\left(\!\frac{1}{2} \tan^{-1}\!\left(\!\frac{\Delta}{h}\!\right)\!\right).
\eea
Taking the number operator of the pair $ \hat{N}_A=a^\dag a+ b^\dag b$, we have
\bea&
\left\langle e^{\lambda  \hat{N}}\right\rangle =\left\langle \left(1+(z-1) a^\dag a\right)\left(1+(z-1) b^\dag b\right)\right\rangle =\nn & \left(1-\left\langle a^\dag a\right\rangle -\left\langle b^\dag b\right\rangle +\left\langle a^\dag a b^\dag b\right\rangle \right)+\nn & z \left(\left\langle a^\dag a\right\rangle +\left\langle b^\dag  b\right\rangle -2 \left\langle a^\dag a b^\dag b\right\rangle \right)+z^2 \left\langle a^\dag a b^\dag b\right\rangle,
\eea
where we denoted $z=e^{\lambda}$.
Substituting $a=u c-v d$ and $b^\dag=v c+u d$ and computing the resulting thermal correlations, which are just free fermions in terms of the $c$ and $d$ operators, we find:
\bea&
\chi =\left(v^2 n_c+u^2 n_d-n_c n_d\right)+z
   \left(1-n_d-n_c+2 n_d n_c\right)+\nn & z^2
   \left(v^2 n_d+u^2 n_c-n_d n_c\right),
   \eea
with $n_d,n_c$ the fermi functions for $c,d$, i.e.:
\bea & n_d=\frac{1}{1+e^{-\beta 
\mathcal{E} }}~~;~~n_c=\frac{1}{1+e^{\beta \mathcal{E} }}.
\eea
After substituting $z$ and  $\mathcal{E} $, and some tedious algebra we recover the result \eqref{counting stat for one pair}.

\section{Overlaps of paired states and counting statistics of charge}
Here we show how to compute the overlap between different BCS like states built out of the vacuum. Such a state is built out of application of pair creation to the vacuum. We will write a (un-normalized) state of this type in the form:
\begin{eqnarray}
|D \rangle=e^{D_{ij}a^{\dag}_i a^{\dag}_j} |0\rangle,
\end{eqnarray}
where the state 
$|0\rangle$  is the vacuum state, so that $a_i|0\rangle=0$ for all $i$, and $D$ is an $n\times n$ antisymmetric matrix. We now derive the following formulas:
\begin{eqnarray}\label{overlap}
\langle D' | D\rangle=(-1)^n {\rm Pf}
\left(
\begin{array}{cc}
  D'^{\dag}   &  -I_n \\
 I_n &   D    
\end{array}
\right)~,
\end{eqnarray}
and as a corollary:
\begin{eqnarray}\label{counting in vacuum with pairing}
\langle e^{i \lambda N_A}\rangle_D={(-1)^n\over \det(1+D^{\dag}D)}  {\rm Pf}
\left(
\begin{array}{cc}
  D^{\dag}   &  -I_n \\
 I_n &   e^{i\lambda P_A} D   e^{i\lambda P_A}  
\end{array}
\right)~.
\end{eqnarray}
To derive these relations, we first note that $|0\rangle$ is the ground state of the Hamiltonian $H=\hat{N}=\sum_i a^{\dag}_i a_i$. We can therefore write the overlap in the following way:
\begin{eqnarray}
\langle D' | D\rangle=lim_{\beta\rightarrow\infty} {1\over Z} {\rm Tr}   [ e^{D_{ij}a^{\dag}_i a^{\dag}_j} e^{-\beta \hat{N}} (e^{D'_{ij}a^{\dag}_i a^{\dag}_j} )^{\dag}]~.
\end{eqnarray}
Going to the Majorana representation we have:
\begin{eqnarray}&
D_{ij}a^{\dag}_i a^{\dag}_j={1\over 8}[D\otimes(\sigma_z-i \sigma_x)]_{\alpha\beta}c_\alpha c_\beta ,\\ &
D_{ij}a_i a_j={1\over 8}[D\otimes(\sigma_z+i \sigma_x)]_{\alpha\beta}c_\alpha c_\beta , \\ & \Sigma _ia_i^{\dagger }a_i=-\frac{1}{4}I_n\otimes \sigma
   _y+\frac{1}{2} {\rm{Tr}}~  I_n.
\end{eqnarray}
We have seen above \eqref{general product of exp majorana}, that:
\begin{eqnarray}
\frac{1}{Z} {\rm{Tr}}~  e^{-\beta  N}\left(e^{D^{'}_{{ij}}a_i^{\dagger
   }a_j^{\dagger }}\right){}^{\dagger }e^{D_{{ij}}a_i^{\dagger
   }a_j^{\dagger }}= \text{   }\sqrt{\det \left(1-n_{\beta
   }+n_{\beta }e^{\frac{1}{2}\left(D'^{\dagger }\otimes \left(\sigma
   _z+\text{i$\sigma $}_x\right)\right)}e^{\frac{1}{2}\left(D\otimes
   \left(\sigma _z-\text{i$\sigma $}_x\right)\right)}\right)}.
\end{eqnarray}
To proceed we make several observations. 

1. Notice that $\left(\sigma _z\pm \text{i$\sigma $}_x\right){}^2=0$ are nilpotent, allowing us to write:
\begin{eqnarray}
e^{\frac{1}{2}\left(D'^{\dagger }\otimes \left(\sigma _z+\text{i$\sigma
   $}_x\right)\right)}e^{\frac{1}{2}\left(D\otimes \left(\sigma
   _z-\text{i$\sigma
   $}_x\right)\right)}=\left(1+\frac{1}{2}\left(D'^{\dagger }\otimes
   \left(\sigma _z+\text{i$\sigma
   $}_x\right)\right)\right)\left(1+\frac{1}{2}\left(D\otimes
   \left(\sigma _z-\text{i$\sigma $}_x\right)\right)\right).
\end{eqnarray}

2. Using the limit:
\begin{eqnarray}
\lim _{\beta \to \infty }n_{\beta }=\lim _{\beta \to \infty
   }\frac{1}{1+e^{4\beta  \left(-\frac{1}{4}I_n\otimes \sigma
   _y\right)}}=\frac{1}{2}\left(I_{2n}+I_n\otimes \sigma
   _y\right)=\frac{1}{2}I_n\otimes \left(1+\sigma _y\right)\equiv P_+~,
\end{eqnarray}
we have
\begin{eqnarray}
\langle D|D'\rangle =\text{   }\sqrt{\det
   \left(P_++P_+\left(1+\frac{1}{2}\left(D'^{\dagger }\otimes
   \left(\sigma _z+\text{i$\sigma
   $}_x\right)\right)\right)\left(1+\frac{1}{2}\left(D\otimes
   \left(\sigma _z-\text{i$\sigma $}_x\right)\right)\right)P_+\right)}.
\end{eqnarray}

3. Finally, we use:
\begin{eqnarray}
P_+\left(\sigma _z\pm \text{i$\sigma $}_x\right)P_+=0,
\end{eqnarray}
and
\begin{eqnarray}
P_+ \left(\sigma _z+\text{i$\sigma $}_x\right)\left(\sigma
   _z-\text{i$\sigma $}_x\right)P_+=2I_n\otimes \left(1+\sigma
   _y\right)=4P_+~.
\end{eqnarray}

Combining the above observations we find:
\begin{eqnarray}
\langle D|D'\rangle =\sqrt{\det
   \left(P_+\left(1+D'^{\dagger }D\otimes I_2\right)P_+\right)}=\sqrt{\det{}_n
   (1+D'^{\dagger }D)}.
\end{eqnarray}
Using the rules of determinants of block matrices we can also rewrite the last expression as:
\begin{eqnarray}
\langle D|D'\rangle =\sqrt{\det  \left(
\begin{array}{cc}
 D'^{\dagger } & -I \\
 I & D \\
\end{array}
\right)}~.
\end{eqnarray}
We now finally have a determinant of an anti-symmetric matrix, and the identity $Pf(A)^2=det(A)$ (valid for anti-symmetric matrices, but not true for general matrices) can be safely used, to get:
\begin{eqnarray}
\langle D|D'\rangle = (-1)^n {\rm Pf}  \left(
\begin{array}{cc}
 D'^{\dagger } & -I \\
 I & D \\
\end{array}
\right)~,
\end{eqnarray}
where the phase $(-1)^n$ was added by demanding $\langle 0|0\rangle=1$ and using:
\begin{eqnarray}
{\rm Pf}  \left(
\begin{array}{cc}
0 & -I_n \\
 I_n & 0 \\
\end{array}
\right)=(-1)^n.
\end{eqnarray}
which establishes Eq. \eqref{overlap}.

Finally, to get equation \eqref{counting in vacuum with pairing}, we write:
\begin{eqnarray}\label{derivation of fcs}
\langle e^{i \lambda N_A}\rangle_D={1\over \langle D|D\rangle} \langle D|e^{i \lambda N_A}|D\rangle={1\over \langle D|D\rangle} \langle D|e^{i \lambda P_A}De^{i \lambda P_A}\rangle~,
\end{eqnarray}
where the last equation is a consequence of 
\begin{eqnarray}
e^{i \lambda N_A}a_i^{\dag} e^{-i \lambda N_A}=\Big\{
\begin{array}{cc}
e^{i \lambda}a^{\dag}_i & i\in A \\
 a^{\dag}_i  & i\notin A \\
\end{array}.
\end{eqnarray}
Now using  \eqref{overlap} in \eqref{derivation of fcs} gives  \eqref{counting in vacuum with pairing}.

\section{Parity fluctuations}
Consider the distribution of the parity operator applied to a subset $A$ of modes of the (complex-fermions) $a_{i}$. In systems with pairing, such parity measurements may be associated with topological effects, and are affected by the presence of majorana zero modes, see, e.g. \cite{burnell2013measuring}.  The parity operator can be written as:
\begin{eqnarray}
\mathcal{P}=(-1)^{\hat{N}_A}=e^{\pi  i
   \hat{N}_A}~.
\end{eqnarray}
Noting that $\langle \mathcal{P}\rangle =p_{\text{e}}-p_{\text{o}}=2 p_{\text{e}}-1$, where
$p_{\text{e}},p_{\text{o}}$ is the probability that the parity of the fermion number is even/odd. 
It's variance is given by:
\bea
\sigma^{2}=\langle \mathcal{P}^{2}\rangle -\langle \mathcal{P}\rangle^{2} =1-\langle \mathcal{P}\rangle^{2} .
\eea
We immediately see that:
\bea
\sigma^{2}=1-\langle \mathcal{P}\rangle^{2} =1-\det \left(1-n_{\beta}+n_{\beta}e^{-i \pi  \sigma
   _yP_A}\right)
\eea
This expression can be further simplified using:
\bea
e^{-i \pi  \sigma _yP_A}={\bf I}_{{2n}}-2P_A\otimes {\bf I}_{2}.
\eea
to get
\bea
\sigma^{2}=1-\det \left(1-2 n_{\beta}P_A\otimes {\bf I}_{2})\right).
\eea
The expectation value of the parity itself can be computed for states such as $|D\rangle$ in \eqref{counting in vacuum with pairing} by simply plugging in $\lambda=\pi$. One can also derive Pfaffian forms for the case not covered in \eqref{counting in vacuum with pairing}, however here we will proceed in the simplest way using the determinant formulas above.

We can write it down in the following way.
 If $\hat{N}_{A}=\sum_{m\in A}\hat{N}_{m}$ for some set of modes, with $\hat{N}_{m}$ is the number operator associated with mode $m$, then we use the relation
\bea
e^{i\pi \hat{N}_{m}}=1-2 \hat{N}_{m}=(1-2 \partial_{\lambda})e^{\lambda\hat{N}_{m}}|_{\lambda=0}~,
\eea
to write
\bea
\langle\mathcal{P}\rangle={1\over Z}{\rm Tr} ~e^{-\beta H_{ij}c_{k}c_{l}}\prod_{m\in A}e^{i\pi \hat{N}_{m}}={1\over Z}\prod_{m \in A}(1-2{\partial\over\partial \lambda_{i}}){\rm Tr}~ e^{-\beta H_{ij}c_{k}c_{l}}\prod_{k\in  A}e^{ \lambda_{k} \hat{N}_{k}}|_{\{\lambda\}=0}.
\eea
We can finally write this expression as:
\begin{eqnarray}&\label{parity der}
\langle\mathcal{P}\rangle=\prod_{m \in A}(1-2{\partial\over\partial \lambda_{m}})e^{{1\over 2}{\sum_{i\in A}\lambda_{i}}}\text{det}^{1/2}
   \left(n_\beta+\left(1-n_\beta\right) e^{\sum_{i\in A}\lambda_{i}\sigma_{y}P_{i}}\right)|_{\{\lambda\}=0}.
\eea
Note that there is no sign ambiguity in this last expression since for Hermitian $H$ and real $\lambda$, ${\rm Tr}~ e^{-\beta H_{ij}c_{k}c_{l}}\prod_{m\in  A}e^{ \lambda_{i} \hat{N}_{m}}$ should always be positive.
In topological applications one considers Majorana zero modes as the simplest known example of non-abelian particles,  $P_A$ can be taken to be the rank one projection operator on the Dirac fermion mode consisting of the two unpaired Majoranas, and thus their ``state'' is determined by the parity. For this type of applications the formula above works very well, since we deal with a particular mode or two, and one can analytically carry out the derivatives above.

\section{Final remarks}
In this paper we have derived formulas for the traces of exponentials of fermion bilinears which include pairing terms, and are not necessarily hermitian. 
We are not aware of previous appearance in the literature of the ``sign ambiguity free'' Pfaffian formulas Eq. \eqref{Pfaff eq} and Eq. \eqref{counting in vacuum with pairing}, nor of a general presentation of Eq. \eqref{general product of exp majorana} and Eq. \eqref{general expectation Majorana}, rather than treatment of special cases.
\begin{comment} It would be of great interest of find a generalization of this expression to several exponents. For small enough $A,B$, where the BCH $C[A,B]$ holds as a series of commutators we can write:
$
 {\rm{Tr}}~  ~ e^{A_{ij}c_i c_j}e^{B_{ij}c_i c_j}=\frac{{\rm Pf} \left(e^{-4 B}e^{-4 A}-e^{4 A}e^{4 B}\right)}{{\rm Pf} \left(e^{-{1\over 2} C[4A,4B]}-e^{ {1\over 2}C[4A,4B]}\right)}~.
$
however, the calculation of $C[4A,4B]$ involves either summing a series or taking the correct square root of the matrix $e^{4 A}e^{4 B}$ which is analytically connected to the identity as function of $A,B$. It would be desirable to have a simpler and straightforward formula avoiding taking such square roots.
\end{comment}

We believe that the (perhaps more practically useful) expressions such as  Eq. \eqref{general product of exp majorana} and Eq. \eqref{general expectation Majorana} may have appeared in various forms in dealing with concrete problems, however we feel it is useful to give them a general  framework and a simple proof, and make them available for other types of problems. Indeed, the above expressions can be straightforwardly applied to numerical and analytical investigations of time dependent problems involving fermions, such as the extension of the study of quasi-particle modeling of X-ray absorption in the cuprates \cite{benjamin2013microscopic,benjamin2014single} to take into account the presence of pairing terms \cite{inprogressDemleretal}.

{\bf Acknowledgement}

I am grateful to Eugene Demler and David Benjamin for discussions, and to Gian Michele Graf for reference \cite{lieb1968theorem}. The work was supported by the NSF CAREER grant DMR-0956053.

%\printbibliography

%\bibliographystyle{abbrv}
\bibliography{/Users/iklich/dropbox/Work/KlichBib.bib}
\end{document}